\documentclass{pasj01}

\usepackage{amsfonts}
\usepackage{graphicx}
\usepackage{amsbsy}
\usepackage{lineno}

\newcommand{\argmax}{\mathop{\rm arg~max}\limits}

\Received{$\langle$reception date$\rangle$}
\Accepted{$\langle$acception date$\rangle$}
\Published{$\langle$publication date$\rangle$}

\begin{document}

\title{Deep-learning Real/Bogus classification for the Tomo-e Gozen transient survey}
\author{
Ichiro \textsc{Takahashi}\altaffilmark{1,*},
Ryo \textsc{Hamasaki}\altaffilmark{2},
Naonori \textsc{Ueda}\altaffilmark{3},
Masaomi \textsc{Tanaka}\altaffilmark{1,4,5},
Nozomu \textsc{Tominaga}\altaffilmark{6,7,2,5},
Shigeyuki \textsc{Sako}\altaffilmark{8,9,10},
Ryou \textsc{Ohsawa}\altaffilmark{8},
and
Naoki \textsc{Yoshida}\altaffilmark{11,5,12}
}%
\altaffiltext{1}{Astronomical Institute, Tohoku University, Sendai, Miyagi 980-8578, Japan}
\altaffiltext{2}{Department of Physics, Faculty of Science and Engineering, Konan University, 8-9-1 Okamoto, Kobe, Hyogo 658-8501, Japan}
\altaffiltext{3}{NTT Communication Science Laboratories, 2-4 Hikaridai, Seika-cho, Keihanna Science City, Kyoto 619-0237, Japan}
\altaffiltext{4}{Division for the Establishment of Frontier Sciences, Organization for Advanced Studies, Tohoku University, Sendai, Miyagi 980-8577, Japan}
\altaffiltext{5}{Kavli Institute for the Physics and Mathematics of the Universe (WPI), The University of Tokyo, 5-1-5 Kashiwanoha, Kashiwa, Chiba 277-8583, Japan}
\altaffiltext{6}{National Astronomical Observatory of Japan, 2-21-1 Osawa, Mitaka, Tokyo 181-8588, Japan}
\altaffiltext{7}{Department of Astronomical Science, School of Physical Sciences, The Graduate University of Advanced Studies (SOKENDAI) 2-21-1 Osawa, Mitaka, Tokyo 181-8588, Japan}
\altaffiltext{8}{Institute of Astronomy, Graduate School of Science, The University of Tokyo, 2-21-1 Osawa, Mitaka, Tokyo 181-0015, Japan}
\altaffiltext{9}{UTokyo Organization for Planetary Space Science, The University of Tokyo, Hongo, Bunkyo-ku, Tokyo 113-0033, Japan}
\altaffiltext{10}{Collaborative Research Organization for Space Science and Technology, The University of Tokyo, Hongo, Bunkyo-ku, Tokyo 113-0033, Japan}
\altaffiltext{11}{Department of Physics, Graduate School of Science, The University of Tokyo, 7-3-1 Hongo, Bunkyo-ku, Tokyo 113-0033, Japan}
\altaffiltext{12}{Institute for Physics of Intelligence, The University of Tokyo, 7-3-1 Hongo, Bunkyo-ku, Tokyo 113-0033, Japan}
\email{ichiro.takahashi@astr.tohoku.ac.jp}

\KeyWords{supernovae: general --- methods: statistical --- surveys}

\maketitle

\begin{abstract}
We present a deep neural network Real/Bogus classifier that improves classification performance in the Tomo-e Gozen transient survey by handling label errors in the training data.
In the wide-field, high-frequency transient survey with Tomo-e Gozen, the performance of conventional convolutional neural network classifier is not sufficient as about $10^6$ bogus detections appear every night.
In need of a better classifier, we have developed a new two-stage training method.
In this training method, label errors in the training data are first detected by normal supervised learning classification, and then they are unlabeled and used for training of semi-supervised learning.
For actual observed data, the classifier with this method achieves an area under the curve (AUC) of 0.9998 and a false positive rate (FPR) of 0.0002 at true positive rate (TPR) of 0.9.
This training method saves relabeling effort by humans and works better on training data with a high fraction of label errors.
By implementing the developed classifier in the Tomo-e Gozen pipeline, the number of transient candidates was reduced to $\sim$40 objects per night, which is $\sim$1/130 of the previous version, while maintaining the recovery rate of real transients.
This enables more efficient selection of targets for follow-up observations.
\end{abstract}
\section{Introduction}\label{sec:intro}
Time-domain astronomy has become an active area in the modern astronomy. Studies of transient phenomena such as supernovae have been rapidly developing in recent years.
To observe transients efficiently, transient surveys have become more wide-field, more sensitive, and more frequent.
As a result, the number of discovered transients has dramatically increased; the reported number of transients reach tens of thousands per year.
In the near future, hundreds of transient objects will be discovered every night with e.g., the Vera C. Rubin Observatory (\cite{2019ApJ...873..111I}).

To detect transients from large data, most of transient surveys implement image subtraction method.
It detects transients by subtracting the past reference image from the observed new image.
By image subtraction, only objects that change brightness, such as transients, can be extracted.
The subtraction method can efficiently detect transients blended in galaxies.
However, this method also has disadvantage, that is, it tends to generate a large amount of fake detections (hereafter called Bogus).
Therefore, development of efficient methods to remove a large number of Bogus has become important.
In order to select the target for follow-up observations, it is necessary to extract the real transients (hereafter called Real) from the detected candidates that include Bogus.
However, with the increase of the scale of observations, the number of Bogus has increased to a level that is not feasible for the human eyes to check.
For example, in Palomar Transient Factory (PTF; \cite{2009PASP..121.1395L}), an order of $10^6$ potential candidates are detected per night (\cite{2013MNRAS.435.1047B}).
Among these, the number of Bogus is estimated to be more than 1000 times greater than that of Real (\cite{2019PASP..131c8002M}).
Thus, conventional selection methods, such as the parameter cutting, are no longer able to narrow down the candidates.
Machine learning techniques are therefore gaining attention as an alternative method.
In the case of Real/Bogus classification, by allowing the machine to learn the relationship between the data of detected objects and their classification results, the machine can classify transient candidates.
If trained in advance, classification is fast and can be performed in real time for a large amount of data.
Various methods for Real/Bogus classification by machine learning have been proposed and implemented in many transient surveys.
In the early era, classification was performed by inputting features extracted from images into Random Forest or Neural Networks (e.g., \cite{2012PASP..124.1175B}; \cite{2013MNRAS.435.1047B}; \cite{2015MNRAS.449..451W}; \cite{morii16}).
Recently, the use of Convolutional Neural Networks (CNN), in which image data are directly input and the machine itself learns features, has become mainstream (e.g., \cite{2017MNRAS.472.3101G}; \cite{2020MNRAS.497.2641T}; \cite{2021MNRAS.503.4838K}; \cite{2021ExA...tmp...61H}).
For example, In the Zwicky Transient Facility (ZTF; \cite{2019PASP..131a8002B}) survey, a CNN-based classifier, braai (\cite{2019MNRAS.489.3582D}) is applied.

The Tomo-e Gozen transient survey is a time-domain survey project, which utilizes a wide-field Tomo-e Gozen camera with 84 CMOS sensors covering a field of view of 20 deg$^2$ per exposure (\cite{2018SPIE10702E..0JS}).
The transient survey is performed with a high cadence of about 3-4 times per night with a typical sensitivity of 18 mag without filters.
The survey is observing at a rate of $10^5$ images/day, and as many as $10^6$ transient candidates are detected every night.
Although the CNN classifier was used to sort out real transients from these candidates, which are mostly Bogus, the classification performance was not sufficient.
There was still a large amount of false positives (an order of $10^3$/day), i.e., Bogus classified as Real.
Thus, we needed a new classifier with a higher performance.

To obtain a higher performance in classification, one can use more complex models.
In general, training of complex models requires large amounts of training data.
In such cases, the training is usually done by using simulated data rather than real data.
However, with millions of samples of simulated data, it becomes infeasible to check their quality of simulation data manually.
As a result, the training data can be contaminated by label errors, e.g., Real mislabeled as Bogus.
When label errors are included in the training data, they cause an adverse impact on performance (e.g., \cite{2022arXiv220309908A}).

This paper describes improvements in the Real/Bogus classification of the Tomo-e Gozen transient survey by using complex machine learning model and by handling label errors.
The structure of this paper is as follows. 
We first introduce the observational dataset used in this work in section 2.
Then, the design of the new classifiers is described in section 3.
The performance of these classifiers is presented in section 4.
In section 5, we discuss key factors for improving the performance and show improvements in the actual operation. 
Finally, we give conclusions in section 6.

\section{Observational data}\label{sec:data}
In this section, we describe our transient survey and dataset used to develop the machine learning classifier. 
We use the optical images from Tomo-e Gozen camera mounted on the 1.05m Kiso Schmidt telescope (\cite{2018SPIE10702E..0JS}).
The Tomo-e Gozen camera is a mosaic camera equipped with 84 CMOS sensors.
Thanks to the fast CMOS readout, the survey data are taken with a rate of 2 frames/sec.
For the transient survey, 12 or 18 consecutive images (6 or 9 sec exposure) are taken with no filter, and these images are stacked.
A typical limiting magnitude of the stacked image is 18 mag.

For transient detection, image subtraction between the stacked image 
and the reference image is performed.
For image subtraction, we use the hotpants software (\cite{2015ascl.soft04004B}), which is based on the method by \citet{1998ApJ...503..325A}.
We use Pan-STARRS1 (PS1) $r$-band data (\cite{2020ApJS..251....4W}; \cite{2020ApJS..251....7F}) as reference images.
Since PS1 data have better sensitivity and better seeing,
the sensitivity of the Tomo-e Gozen image is not degraded even after the image subtraction,
which enables the efficient transient detection.
Also, as the Tomo-e Gozen camera is a newly developed camera, 
there were no deep stacked images at the beginning of the operation.
Thus, the use of the existing PS1 data enables the transient survey 
even at the early stage of the operation.

Thanks to the high observing rate, as many as $10^5$ stacked images
are taken every night in the Tomo-e Gozen transient survey.
The number of transient candidates can reach $10^6$ in one night.
As in other transient surveys,
the detected candidates are by far dominated by Bogus detection.
Also, image subtraction between different telescopes/instruments
tend to cause more Bogus due to the differences in various factors of the data, for example, response functions or pixel scales.
In this paper, we show that this difficulty can be overcome by developing the high-performance, complex machine learning classifier (section \ref{sec:method}).

\begin{figure}[thbp]
  \begin{center}
     \includegraphics[width=\columnwidth]{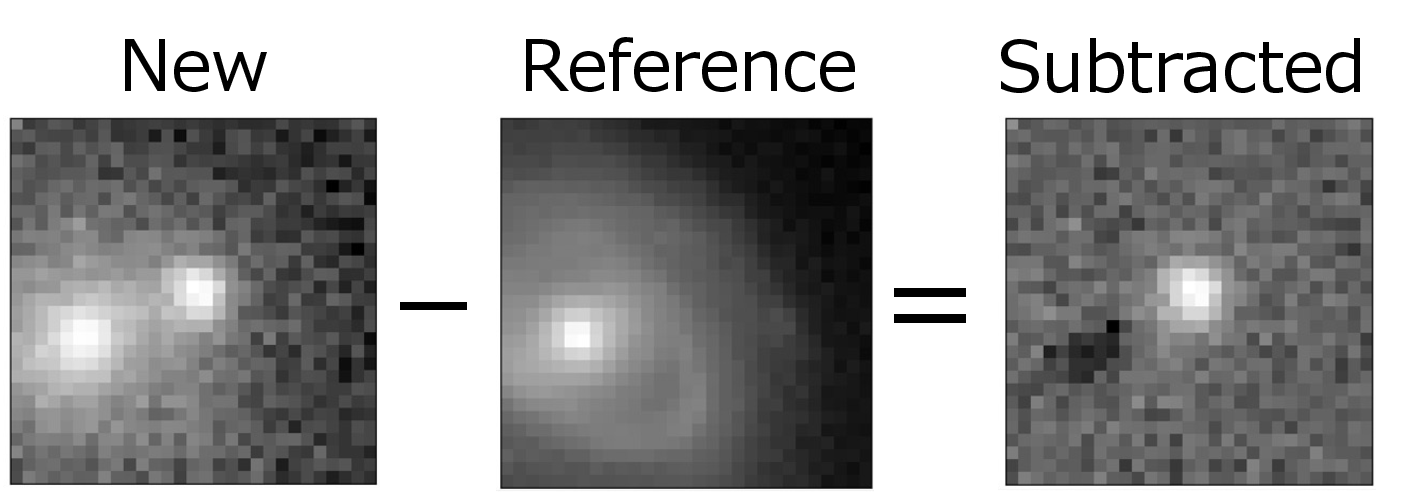}
  \end{center}
  \caption{%
  An example of input images (new, reference, and subtracted images from left to right).
  }%
  \label{fig:imsub}
\end{figure}

We here describe the dataset used to develop the machine learning classifier.
For the Real/Bogus classification, 
we use a set of three images: observed new image, reference image, and subtracted image.
For each, a cutout image around the transient candidate (29 $\times$ 29 pixels)
is used as a input of the classifier.
Figure\ \ref{fig:imsub} shows an example of cutout images.

Since we need large dataset for training of complex machine,
we use artificial objects as Real for training dataset.
For this purpose, we constructed a point spread function (PSF) for each image by measuring the shapes of 
stars in the image, and then embedded the constructed PSF into the observational images 
with various brightness.
The artificial objects were embedded at two kinds of locations:
(1) uniform distribution around galaxies, and
(2) random distribution for the entire region of the images.
Here, (1) and (2) mimic normal transients and hostless transients, respectively.
We prepared about $6\times10^5$ samples for each case.
When embedding artificial sources around galaxies, we randomly selected objects registered as extended sources in the Pan-STARRS catalog.

\begin{figure}[thbp]
  \begin{center}
     \includegraphics[width=\columnwidth]{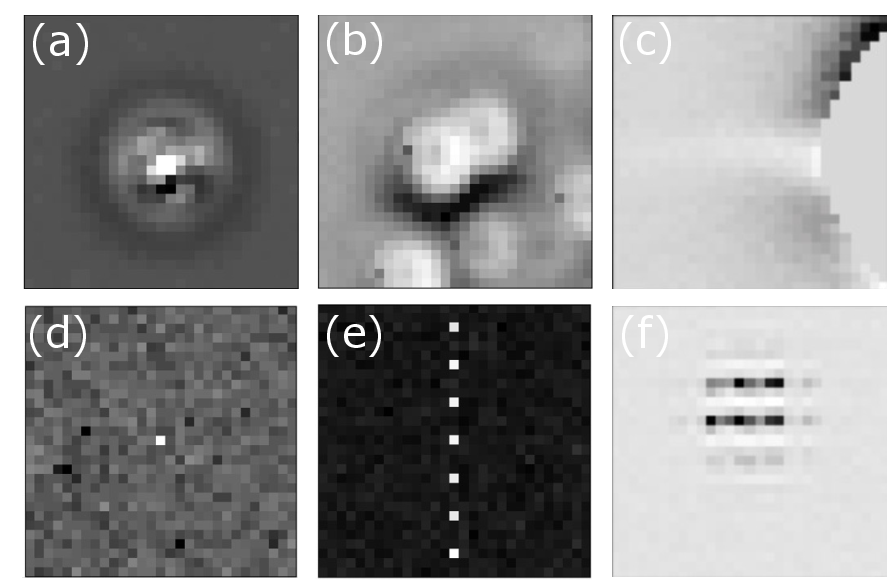}
  \end{center}
  \caption{%
  Examples of Bogus detection.
  }%
  \label{fig:bogus_example}
\end{figure}

For Bogus samples of training data, 
we used actual Bogus objects which were detected in the subtracted images of Tomo-e Gozen. 
Figure\ \ref{fig:bogus_example} shows examples of Bogus in our dataset.
The total number of the Bogus samples is $2\times10^6$. 
The majority of them are false detections due to failed subtraction ((a) and (b) in figure\ \ref{fig:bogus_example}). 
Other cases include the false detection due to the diffracted light from bright stars (c), hot pixels of the sensor (d), and 
artificial noise patterns due to malfunction of the data acquisition system ((e) and (f)).
It is emphasized that a small amount of Real transients can be included in the Bogus samples since we assume that all the detected objects are Bogus (i.e., label error, see section\ \ref{sec:semisupervised}).

For the validation dataset, we used samples extracted from the actual observational data from January to April 2021.
The Real dataset includes 125 objects reported to the Transient Name Server (TNS)\footnote{TNS $\langle$https://www.wis-tns.org$\rangle$.} that were observed by Tomo-e Gozen.
Since some of them were detected multiple times, the total number of Real samples is 363.
As Bogus samples, we used Bogus objects in the same images where the Real objects are detected.
The total number of Bogus is 255,777.
The actual Bogus to Real ratio is much higher than this sample ratio (see section \ref{sec:actualope}).
Nevertheless, this sample ratio was adopted because the size of Bogus dataset would exceed that of the training dataset if we adopt the actual ratio ($\sim$ 1:$10^6$).
The number of samples in training and validation dataset are summarized in table\ \ref{tab:datasetnum}.

Finally, we test our classifiers in the actual operation of the Tomo-e Gozen transient survey (section \ref{sec:actualope}).
For this purpose, we use the data taken for five nights, which correspond to about $5 \times 10^5$ images and $5 \times 10^6$ detections in total.

\begin{table}[htbp]
\tbl{Number of training and validation dataset.}{
\begin{tabular}{lrrrr}
\noalign{\vskip 1mm}
\hline
Dataset  &       & Number  & Note \\
\hline
Training & Real  & 1,224,773 & Artificial \\
         & Bogus & 2,031,193 & Actual \\
\hline
Validation     & Real  & 363    & TNS \\
         & Bogus & 255,777    & Actual \\
\hline
\end{tabular}
}\label{tab:datasetnum}
\end{table}
%

\section{Method}\label{sec:method}
To improve the performance of the conventional classifier, we modify the neural network into the one with a more complex structure (section\ \ref{sec:classifier}).
In addition, in order to take advantage of neural networks with complex structures, we propose a new training method\footnote{The source code of our proposed method is available at $\langle$https://github.com/ichiro-takahashi/tomoe-realbogus$\rangle$.}
devised for objective functions and dataset handling (section\ \ref{sec:training_methods}).

\subsection{Model architecture}\label{sec:classifier}
This subsection describes the model structure of a simple conventional model, which serves as a baseline, and a more complex model we propose.
Hereafter, we call them as ``Simple model'' and ``Complex model'', respectively.
Both models perform binary classification (Real or Bogus) for a detected object by using three input images.
Since Real class is important here, Real class and Bogus class are defined as positive class and negative class, respectively.

\subsubsection{Simple model}
\begin{figure}[thbp]
  \begin{center}
     \includegraphics[width=\columnwidth]{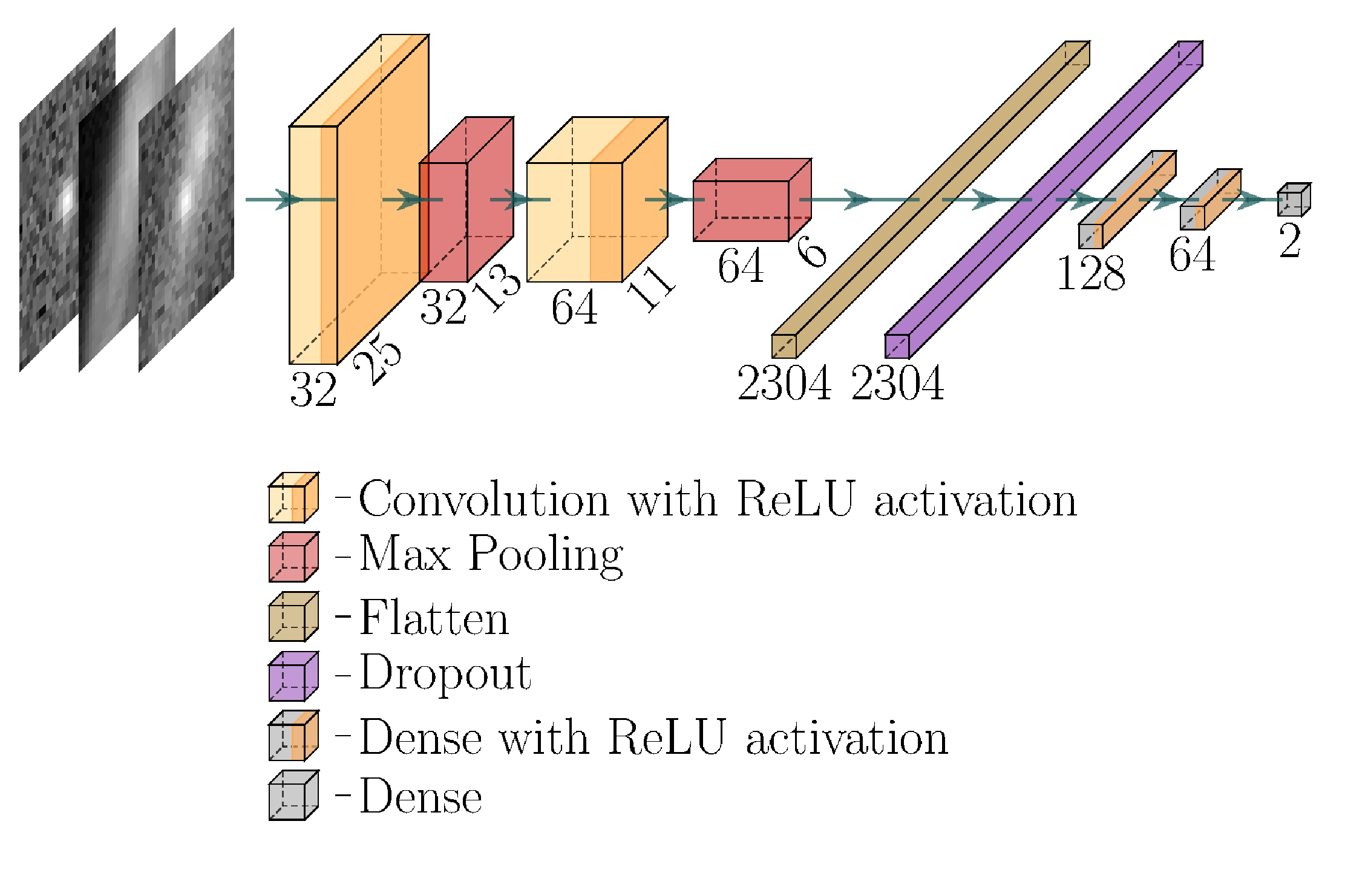}
  \end{center}
  \caption{%
  Architecture diagram of the Simple model.
  Since no padding is performed in the convolution layer, the spatial size of the features decreases with each pass through the convolution layer.
  The model adopts kernel\_size$=$5 for the first convolutional layer and kernel\_size$=$3 for the second convolutional layer.
  The flatten layer collapses the input (6,6,64) features into a vector with a size of 2304.
  In the dropout layer, the probability of the value being 0 is set to 0.3.
  This figure was generated by PlotNeuralNet \footnote{PlotNeuralNet $\langle$https://github.com/HarisIqbal88/PlotNeuralNet$\rangle$.} and modified.
  }%
  \label{fig:simple_CNN}
\end{figure}

As shown in figure\ \ref{fig:simple_CNN}, the Simple model consists of two convolutional layers in the first half and three fully connected layers in the second half.
This structure follows the VGG model (\cite{simonyan2015deep}), which is a basic model structure for image classification tasks using deep learning.
\footnotetext{PlotNeuralNet $\langle$https://github.com/HarisIqbal88/PlotNeuralNet$\rangle$.}
In order to keep the range of values inside the network consistent, we normalize the input images.
The normalization is performed on the original image $\mathbf{u}_{i,c} \in \mathcal{R}^{H \times W}$ of the $c$th channel in the $i$th sample ($H$ and $W$ are the height and width of each image) as follows:
\begin{eqnarray}
    \mathbf{x}_{i,c} = \frac{\mathbf{u}_{i,c} - \min \left( \mathbf{u}_{i,c} \right)}{\max \left( \mathbf{u}_{i,c} \right) - \min \left( \mathbf{u}_{i,c} \right)} ,
\end{eqnarray}
where $\max\left(\right)$ and  $\min\left(\right)$ are functions that return the maximum and minimum values of the image, respectively.
The output of the network is a two-dimensional vector.
By normalizing the output vector with the softmax function, it can be interpreted as a probability that the object is Real.

\subsubsection{Complex model}
\begin{figure*}[htbp]
  \begin{center}
    \includegraphics[width=2\columnwidth]{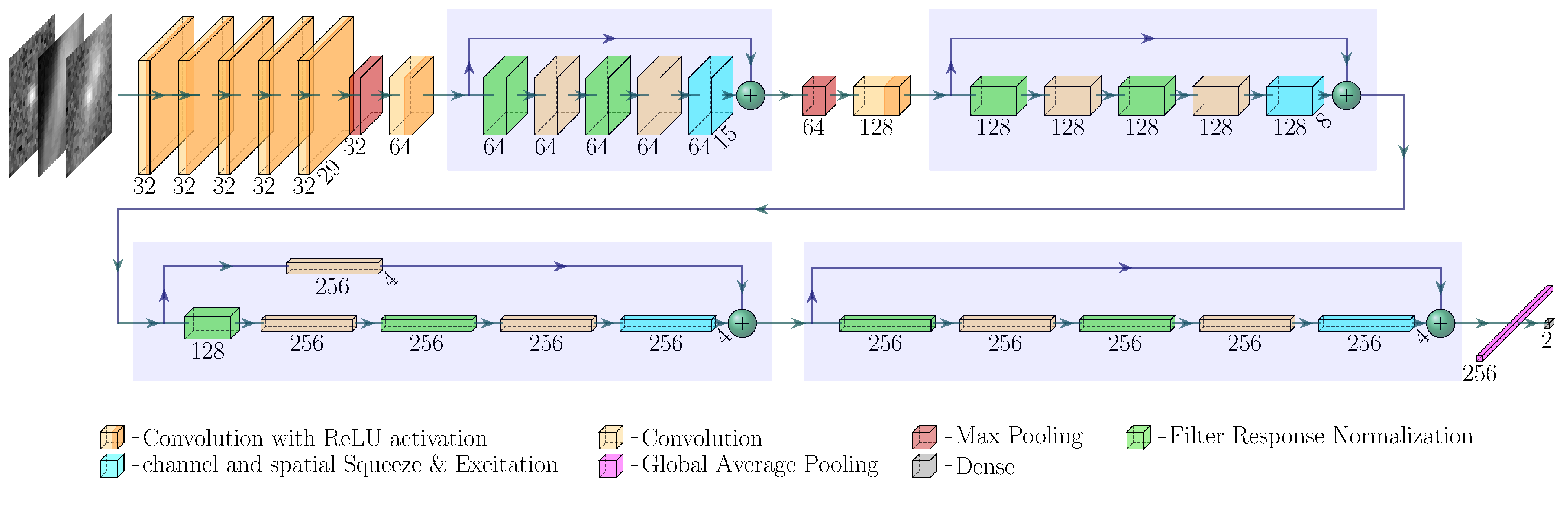}
  \end{center}
  \caption{%
  Architecture diagram of the Complex model.
  In the convolution layer, the kernel size is 5 only in the first layer after input and it is 3 in the remaining layers. Also, padding is performed to make the size in the spatial direction the same for input and output.
  Each region with a light blue background is a Residual Block.
  The third Residual Block from the input side has a convolution layer with stride$=$2 in the skip connection to adjust the feature size.
  The global averaging pooling calculates averages in the spatial direction for each channel and downsamples features of size (4, 4, 256) into a 256-dimensional vector.
  This figure was generated by PlotNeuralNet and modified.
  }%
  \label{fig:new_model}
\end{figure*}

To achieve higher performance than the Simple model, we increase the number of layers in the new Complex model. However, increasing the number of layers makes training more difficult.
One famous model that addresses this problem is ResNet (\cite{7780459}; \cite{10.1007/978-3-319-46493-0_38}), which has various extended versions.
Among many extensions of ResNet, we have adopted SE-ResNet, which includes a channel and spatial Squeeze \& Excitation (csSE) layer (\cite{10.1007/978-3-030-00928-1_48}).
Figure\ \ref{fig:new_model} shows the architecture diagram of the Complex model.
ResNet is a network structure consisting of stacked Residual Blocks, each of which consists of convolutional layers and a skip connection between the input and the output of the block.
The skip connection eliminates the vanishing gradient problem and this facilitates performance of training.
Since it is difficult to learn the identity map only with convolutional layers, there is a problem that too many convolutional layers degrade the performance.
The skip connection makes it easier to learn the identity map in the entire Residual Block. Therefore, high performance can be achieved even in a deep network of stacked Residual Blocks.
Since the csSE layer emphasizes the effective parts of the features for classification, SE-ResNet is expected to improve classification performance compared to the original ResNet.
One component of the Residual Block is the Batch Normalization layer (\cite{10.5555/3045118.3045167}).
It has the effect of stabilizing and speeding up training which contributes to the success in learning deep networks.
However, Batch Normalization has several weaknesses, one of which is its performance degrading when the samples in a batch are highly correlated.
This is a concern when there are many similar images with a high correlation, such as astronomical images.
Therefore, instead of using Batch Normalization, we used Filter Response Normalization (\cite{Singh_2020_CVPR}), which is not affected by the correlation between samples because the normalization is done per channel and per sample.
The input to the network, its normalization method, and the format of the output are the same as for the Simple model. 

\vskip\baselineskip
\subsection{Training methods}
\label{sec:training_methods}
This subsection describes the training methods for the conventional and new classifiers.
We test multiple methods for three phases of training: (1) how to treat the training data, (2) which objective function is used, and (3) how to handle label errors.
\subsubsection{Treatment of training data} \label{sec:treatment_data}
For the treatment of training data, two methods are tested: the first one is to prepare the training data for each CMOS sensor and prepare a classifier for each sensor; the other is to combine the training data of all the sensors and train a single classifier to classify the data of all the sensors.
These tests are performed because it is not obvious which option gives a better performance, i.e, having each classifier specialized for each sensor to care the sensor diversity or having a unified classifier for all the sensors.

\subsubsection{Objective functions}
\label{sec:objective_functions}
For objective functions for training, three types of functions are used:
the cross-entropy function, exp-Cross-hinge function (\cite{misc/27342721}), and local distributional smoothness function (\cite{miyato2016distributional}).
They are used for the purpose of loss function for individual samples, loss function for the entire dataset, and loss function to make the classifier robust to input perturbation, respectively.
Each training image $\mathbf{x}$ prepared in section \ref{sec:data} is paired with a teacher label $y$ indicating whether the image is positive class or negative class.
In this work, we dare to ignore the labels of some of the training data (we call this procedure ``unlabel'').
We describe why we ignore the labels and how to select them from the dataset in section \ref{sec:mislabel_detection}.
The objective function is defined as follows:
\begin{eqnarray}
    L\left( \left\{\mathbf{x}_l, y\right\}, \left\{\mathbf{x}_u\right\} ; \boldsymbol{\theta} \right) 
    & = & \lambda_\mathrm{ce} L_\mathrm{ce} \left( \left\{\mathbf{x}_l, y\right\} ; \boldsymbol{\theta} \right) \nonumber \\
    & & + \lambda_\mathrm{ech} L_\mathrm{ech} \left( \left\{\mathbf{x}_l, y\right\} ; \boldsymbol{\theta} \right) \nonumber \\
    & & + \lambda_\mathrm{lds} L_\mathrm{lds} \left( \left\{\mathbf{x}_l\right\} \cup \left\{\mathbf{x}_u\right\} ; \boldsymbol{\theta} \right) . \label{eq:objective}
\end{eqnarray}
Here, $\left\{\mathbf{x}_l, y\right\}$ is a set of pairs of labeled image and its label, $\left\{\mathbf{x}_u\right\}$ is a set of unlabeled images, $\boldsymbol{\theta}$ is a set of trainable variables of the neural network,
and $L_\mathrm{ce}$, $L_\mathrm{ech}$, and $L_\mathrm{lds}$ are cross-entropy loss, exp-Cross-hinge loss, and local distributional smoothness (LDS) loss, respectively. 
The exp-Cross-hinge function is related to Area Under the Curve (AUC) maximization. The local distribution smoothness function is a key component of the Virtual Adversarial Training (VAT).
The details of each term are described in the Appendix.

Three scalar hyper-parameters $\lambda_\mathrm{ce}$, $\lambda_\mathrm{ech}$, and $\lambda_\mathrm{lds}$ control the effect of each term.
By setting one or two $\lambda$ values in equation (\ref{eq:objective}) to 0, we can create several variations of the objective functions.
The objective function of the Simple model corresponds to the case $\left( \lambda_\mathrm{ce}, \lambda_\mathrm{ech}, \lambda_\mathrm{lds} \right)^T = \left(1, 0, 0\right)^T$, which equals to cross-entropy function.
In section \ref{sec:semisupervised}, we test four patterns of objective functions for training the Complex model, where $\lambda_\mathrm{ce}$ is always non-zero and $\lambda_\mathrm{ech}$ and $\lambda_\mathrm{lds}$ can be zero or non-zero.
It is necessary to tune the none zero element of the hyper-parameter $\left( \lambda_\mathrm{ce}, \lambda_\mathrm{ech}, \lambda_\mathrm{lds} \right)^T$.
This tuning is performed by grid search, and multiple trials are made at each grid point to avoid possible variations in the results due to model initial values and other factors. 
The best combination of $\lambda$ values is the one that produces the model with the highest performance among the results of these trials.
In training, $\boldsymbol{\theta}$ is updated using the stochastic gradient descent to minimize the value of the objective function defined above.

\subsubsection{Handling of label errors}
\label{sec:mislabel_detection}
The training data we prepare are not always perfect. Some of them have incorrect labels or are difficult to label.
It would be better to correct these label errors and train the classifiers with a clean training dataset.
However, for large dataset, it is not practical to manually check and correct all of these samples as it requires an enormous amount of human effort.
Therefore, we split the training into two stages to handle label errors.
In the first stage of training, the machine finds samples that are likely to be mislabeled.
Then, the second stage of training is performed by handling the samples found in the first stage.
In both stages, the Complex model is used.
To find label errors, we first train the classifier with the original training data by utilizing the fact that the ratio of label errors to the training data is sufficiently small.
The classifier then classifies the training data themselves and identifies label errors.
Specifically, the training dataset is divided into five categories: one is used for evaluation and the remaining four are used to train the classifier.
Since there are five different ways to select data for evaluation, all the training data can be evaluated by five training and evaluation cycles in total.
This method is the same as in \citet{northcutt2021confidentlearning}.
After classifying the training data themselves, we determine which samples are likely to have label errors based on the output of the classifier.
Unlike \citet{northcutt2021confidentlearning}, we simply set the classification boundary to a probability of 0.5, and regard all misclassified samples as samples with potential label errors.
Although the threshold value of 0.5 makes more samples to be potential label errors than the value used by \citet{northcutt2021confidentlearning}, the number of potential label errors is small ($\sim1\%$) relative to the entire dataset in our case (section\ \ref{sec:semisupervised}).
Since the samples with potential label errors are used in semi-supervised learning as unlabeled samples, the effect of the overestimation is minor in the second stage of training described below.

In the second stage of training, we try two different methods to handle label errors.
The first method is simply to remove the samples with potential label errors from the training dataset.
The second method is setting the samples with potential label errors as ``unlabeled'' and then performing semi-supervised learning.
The VAT does not use the labels in the computation of the objective function, which allows semi-supervised learning.
Semi-supervised learning avoids the adverse effects of samples with label errors, while effectively utilizing them as training data.

\section{Results}\label{sec:results}
In this section, we summarize the Real/Bogus classification performance for the validation dataset with the various combinations of the models and training methods described in the previous sections.
\subsection{Effects of training data treatments}\label{sec:supervised}
\begin{figure}[thbp]
  \begin{center}
     \includegraphics[width=\columnwidth]{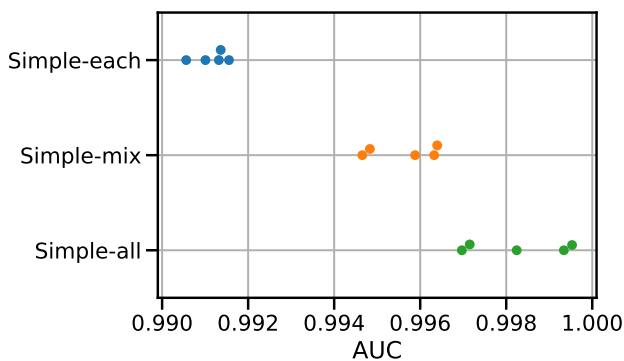}
  \end{center}
  \caption{%
  Comparison of AUC for the Simple models. The five points in each model represent the performance variation with different initial seed values.
  }%
  \label{fig:auc_simple}
\end{figure}

\begin{figure}[thbp]
  \begin{center}
     \includegraphics[width=\columnwidth]{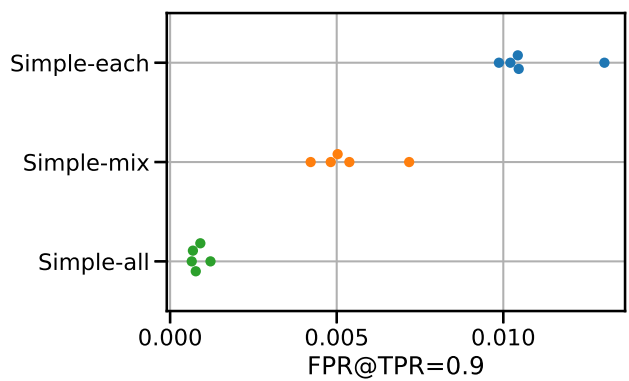}
  \end{center}
  \caption{%
  Same as figure \ref{fig:auc_simple}, but for FPR at TPR=0.9.
  }%
  \label{fig:fpr_simple}
\end{figure}

We compare the performance of the three cases of the Simple models: (1) the model trained for each CMOS senor (Simple-each), (2) the model that has the same total number of samples as (1) but trained with samples of all sensors (Simple-mix), and (3) the model trained with the entire data for all sensors (Simple-all).
First, we examine the impact of the sample diversity from multiple sensors on performance.
The conventional classifier adopts the Simple-each approach and are trained using only cross-entropy loss as the objective function.
Classification based on a dataset from a single sensor can cause overfitting, in which the unique ``habits'' of the dataset are used to classify the data.
By training on datasets from multiple sensors, it is possible to learn more essential features that do not depend on unique ``habits'', and thus to obtain the effect of data augmentation.
Second, we study the effect of the size of the training data.
In the Simple-all case, the total number of actual training data is larger than that of the Simple-each case by a factor of 84 (the number of sensors).
The performance of the classifier can be evaluated by the Receiver Operating Characteristic (ROC) curve.\footnote{ROC curve shows the true positive rate (TPR) and false positive rate (FPR) values measured with different thresholds.
The better the classification performance is, the closer the ROC curve is to the upper left corner of the plot.}
The AUC of the ROC curve indicates the overall performance.
For the Simple-each case, since a classifier is prepared and tested for each sensor, we measure the performance of these cases by combining the results of all sensors.
Figures\ \ref{fig:auc_simple} and \ref{fig:fpr_simple} show the AUC and FPR for each Simple model, respectively.
To see the variation of the results, we plot the results of five training runs with different initial seed values for each case.
The FPR in figure\ \ref{fig:fpr_simple} is defined at a threshold when the TPR is 0.9.
Comparison between the Simple-each case and the Simple-mix case shows that data-mixing with different sensors gives better results thanks to the data augmentation effect for the same size training data.
Furthermore, the Simple-all case achieved better results.
This means that both the larger size of the training data and the mixing of data from multiple sensors contribute to the improvement.
%

\subsection{Effects of label error handling}\label{sec:semisupervised}
\begin{figure*}[htbp]
    \begin{tabular}{cc}
        \begin{minipage}{0.5\hsize}
            \begin{center}
                \includegraphics[width=\columnwidth]{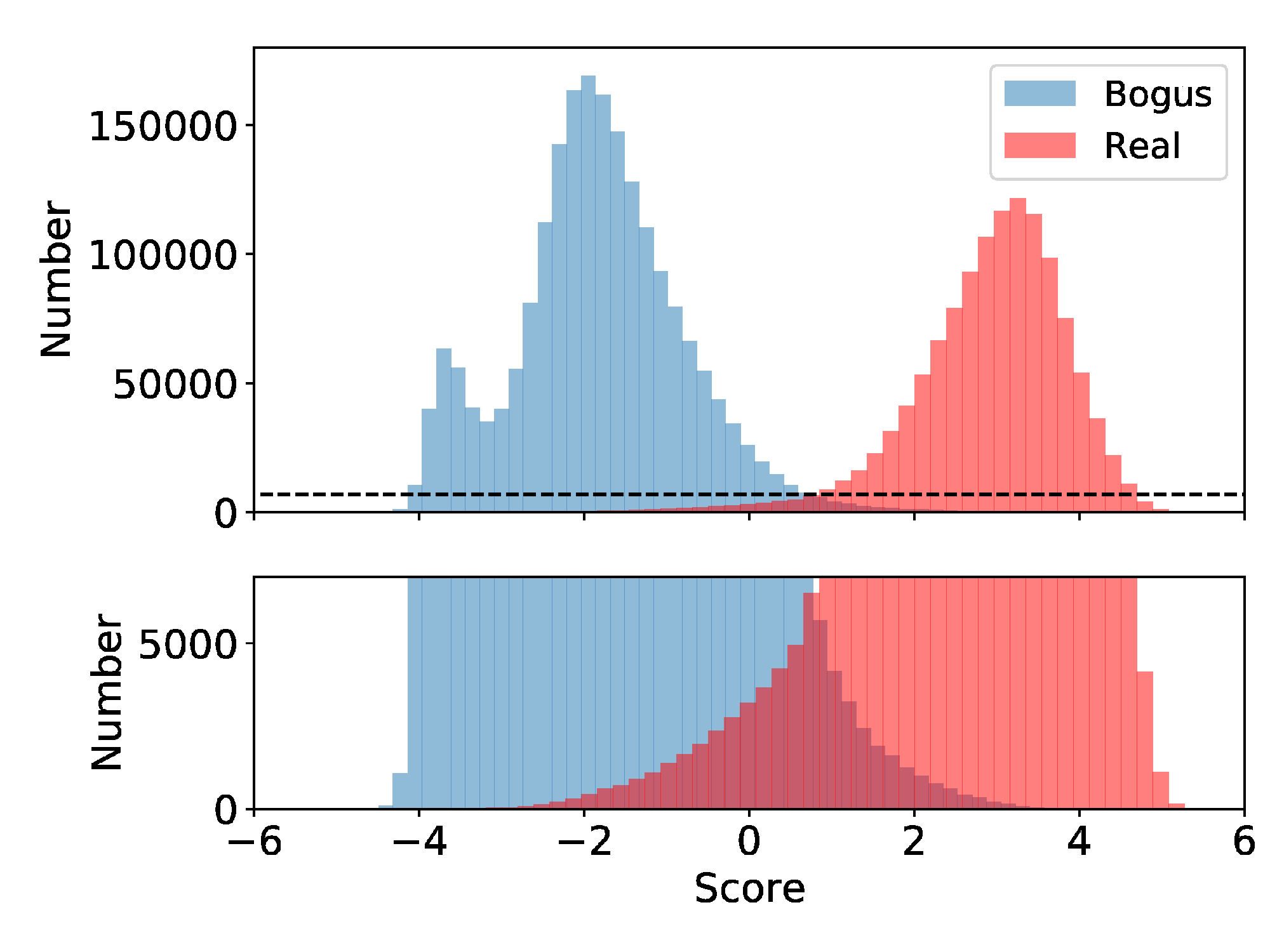}
            \end{center}
        \end{minipage}
        \begin{minipage}{0.5\hsize}
            \begin{center}
                \includegraphics[width=\columnwidth]{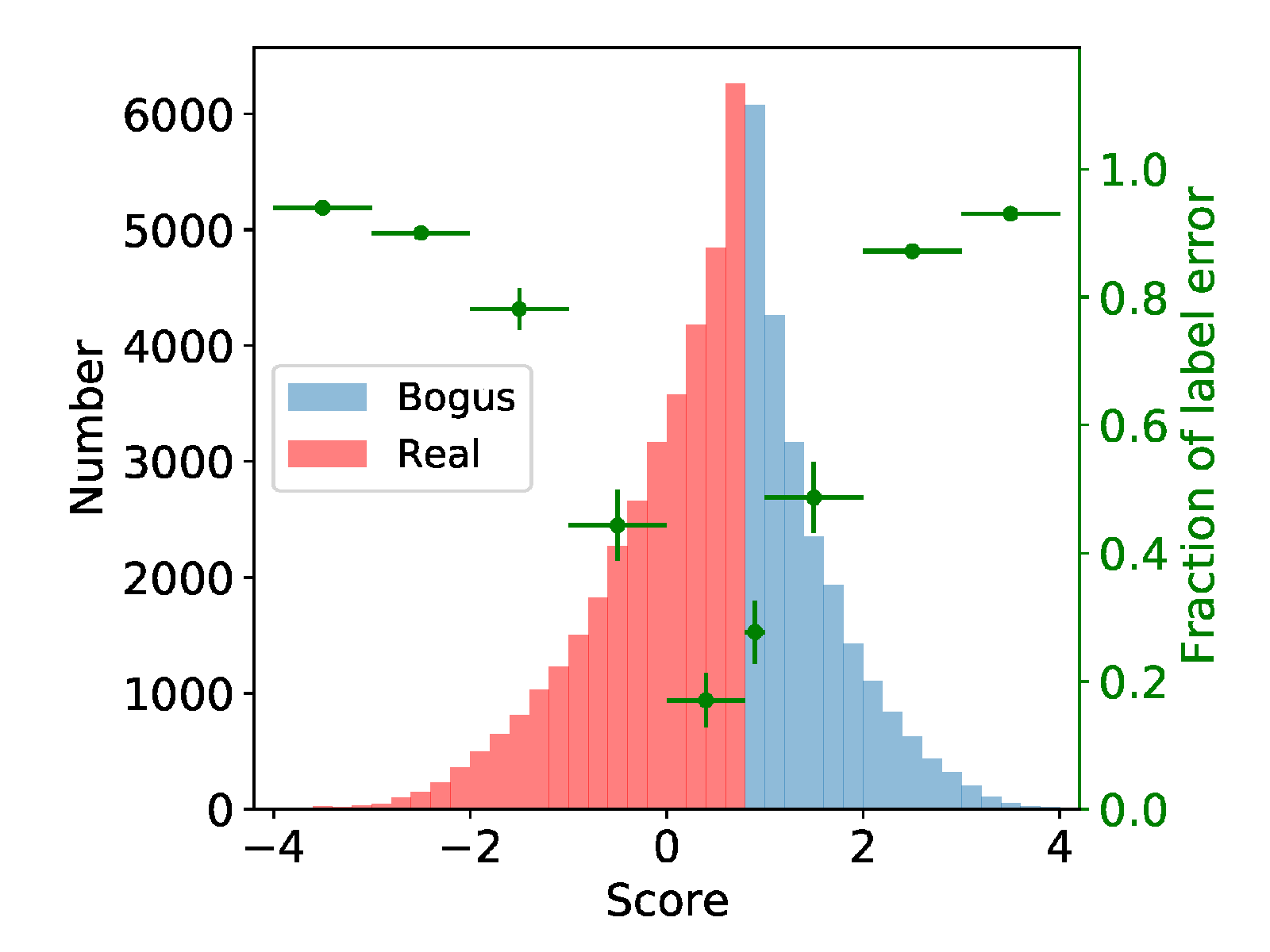}
            \end{center}
        \end{minipage}
    \end{tabular}
    \vspace{2mm}
    \caption{%
    Left: Score distribution of the training dataset. Right: Score distribution (left axis) and fraction of label errors estimated from a sample survey by human eyes (right axis).
    The sample survey is performed at the edges of the score distribution.
    }
    \label{fig:mislabel}
\end{figure*}

We here investigate whether treatment of label errors improves the performance of the Complex model.
For the baseline of the performance comparison, we use the performance of the Simple-all case.
When multiple objective functions are combined, optimal values for the weight of the functions, i.e., $\lambda_\mathrm{ce}$, $\lambda_\mathrm{ech}$, and $\lambda_\mathrm{lds}$, are obtained from the parameter search for each case.
The optimal values in each case are summarized in table\ \ref{tab:performance}.
First of all, we estimate the fraction of label errors in the training data.
For the training data, we use the entire data from all the sensors, as in the Simple-all case.
The left panel of figure\ \ref{fig:mislabel} shows the score distribution when the training data themselves are classified using a Complex model trained with cross-entropy loss for the label error identification.
The score here is defined by the score function in equation (\ref{eq:score}), not probability.
We pay particular attention to samples that the classifier misclassifies, i.e., those in the tail of the Real and Bogus distributions.
For these samples, we check the images by visual inspection.
Since there are a large number of samples even in the tail, we conducted a sample survey to estimate the fraction of label errors.
The fraction is evaluated by visually counting the label errors from randomly selected samples for each score bin.
The right panel of figure\ \ref{fig:mislabel} shows the distribution of the scores and the fraction of label errors in each score bin. 
In fact, we find that the fraction of label errors also increases at the edges of the distributions.
Based on the estimated fractions of label errors, the contamination ratio of label errors in the training data is about 0.6\% for Bogus and 1\% for Real.

\begin{figure}[thbp]
  \begin{center}
     \includegraphics[width=\columnwidth]{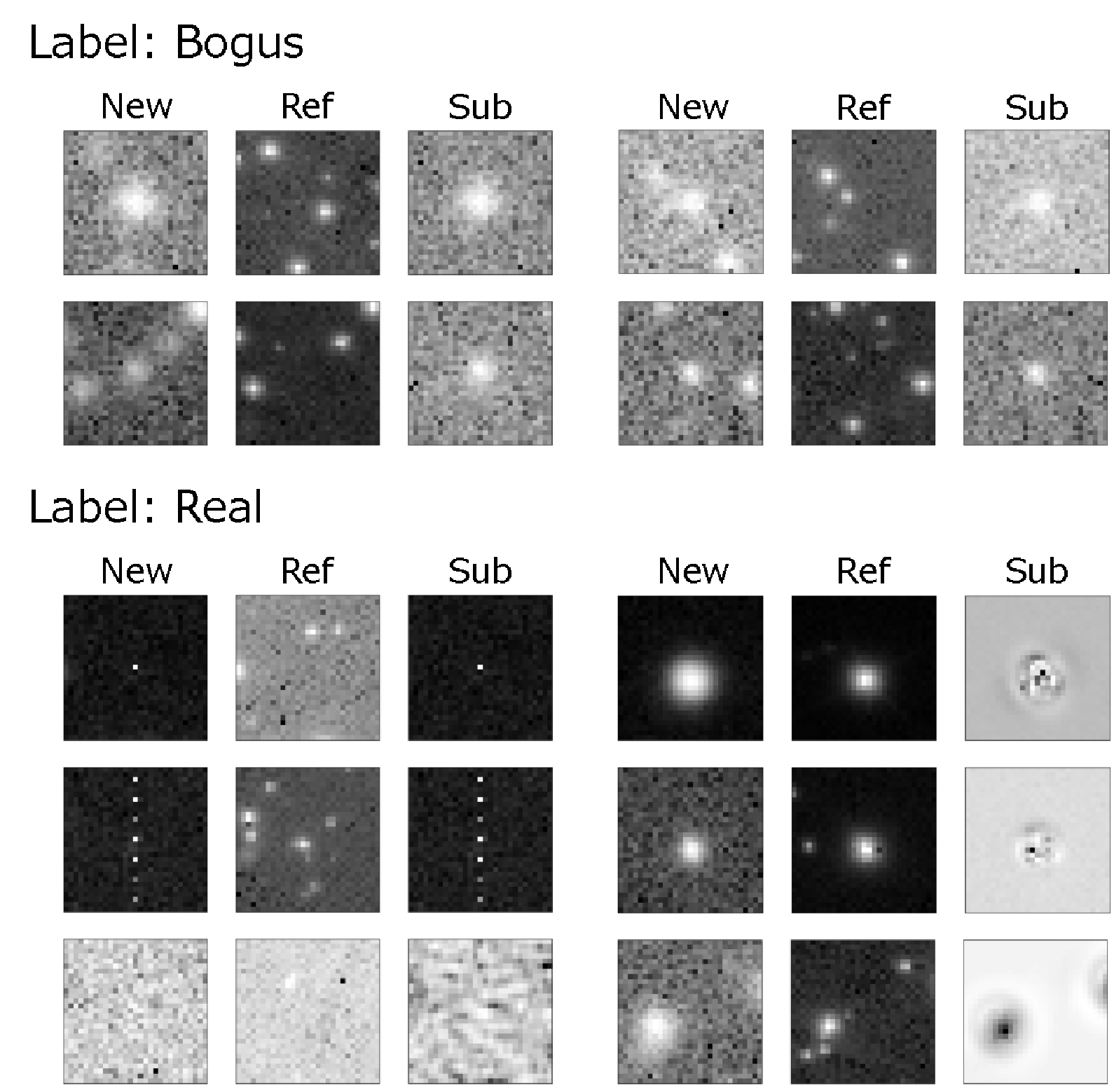}
  \end{center}
  \caption{%
  Examples of the samples with label errors.
  }%
  \label{fig:label_error_examples}
\end{figure}

\begin{figure}[thbp]
  \begin{center}
     \includegraphics[width=\columnwidth]{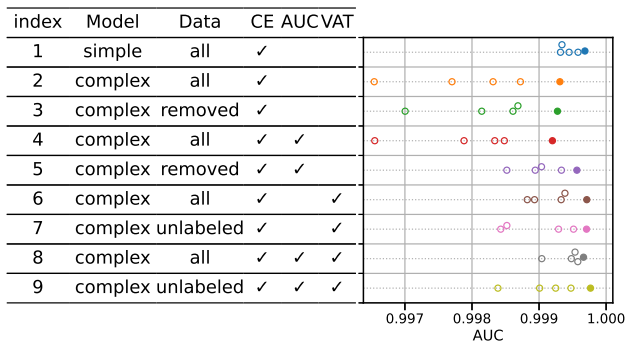}
  \end{center}
  \caption{%
  Summary of AUC for all the classifiers. Five points in each case represent the performance variation with different initial seed values. The filled points indicate the training with the highest AUC for each case. The CE, AUC, and VAT columns indicate whether or not each term of the objective function is used.
  }%
  \label{fig:auc_complex}
\end{figure}

\begin{figure}[thbp]
  \begin{center}
     \includegraphics[width=\columnwidth]{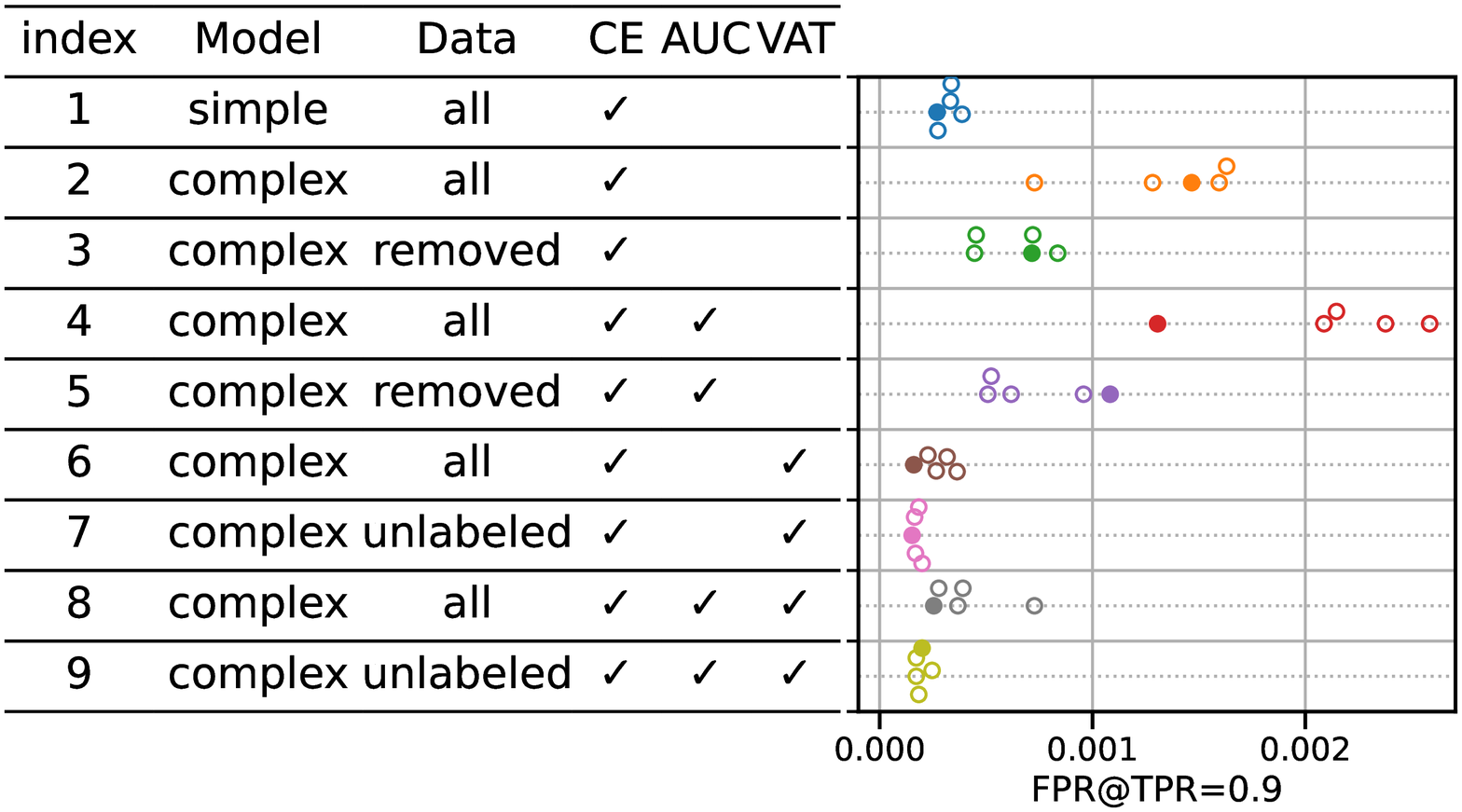}
  \end{center}
  \caption{%
  Same as figure \ref{fig:auc_complex}, but for FPR at TPR=0.9.
  }%
  \label{fig:fpr_complex}
\end{figure}

Figure\ \ref{fig:label_error_examples} shows examples with label errors at the edge of the score distribution.
Among these, those mislabeled as Bogus clearly show objects that appear to be transients.
On the other hand, those mislabeled as Real have embedded artificial stars, but with a distinct Bogus detection.
In other words, the labels are not correct for these samples and the machine actually classifies them correctly.
Then, we investigate the effect to the classification performance for two different ways to handle label errors.
First, we examine the method in which we simply remove the samples with potential label errors. 
The AUC and FPR with potential label errors (indices 2 and 4) and without potential label errors (indices 3 and 5) are shown in figures\ \ref{fig:auc_complex} and \ref{fig:fpr_complex}, respectively.
Two classifiers are used in the comparison: one with CE only and the other with CE+AUC as the objective function.
In figures \ \ref{fig:auc_complex} and \ref{fig:fpr_complex}, the CE, AUC, and VAT columns indicate whether or not $L_\mathrm{ce}$, $L_\mathrm{ech}$, and $L_\mathrm{lds}$ are used, respectively.
Both classifiers perform better when the samples with potential label errors are removed (indices 3 and 5).
The FPR is significantly lower when the samples with potential label errors are removed while the AUC shows no significant difference due to high variation in each seed value. 
However, in all the cases, the classifiers removing potential label errors do not perform better than the Simple-all case.
Next, we examine the semi-supervised learning method in which all samples with potential label errors are unlabeled.
The AUC and FPR for the model with CE+AUC+VAT as the objective function in each case are shown as indices 8 and 9 in figures\ \ref{fig:auc_complex} and \ref{fig:fpr_complex}.
In this comparison, the method that handles potential label errors by unlabeling and by performing semi-supervised learning (index 9) shows better results than the supervised learning with potential label errors remaining (index 8). 
Furthermore, the semi-supervised learning method (index 9) yields a lower FPR at TPR=0.9 than the method that removes potential label errors (indices 3 and 5) and the Simple-all case (index 1).
This means that the semi-supervised learning method can achieve good performance even if the training data contain label errors.

\begin{figure*}[Htbp]
  \begin{center}
     \includegraphics[width=2\columnwidth]{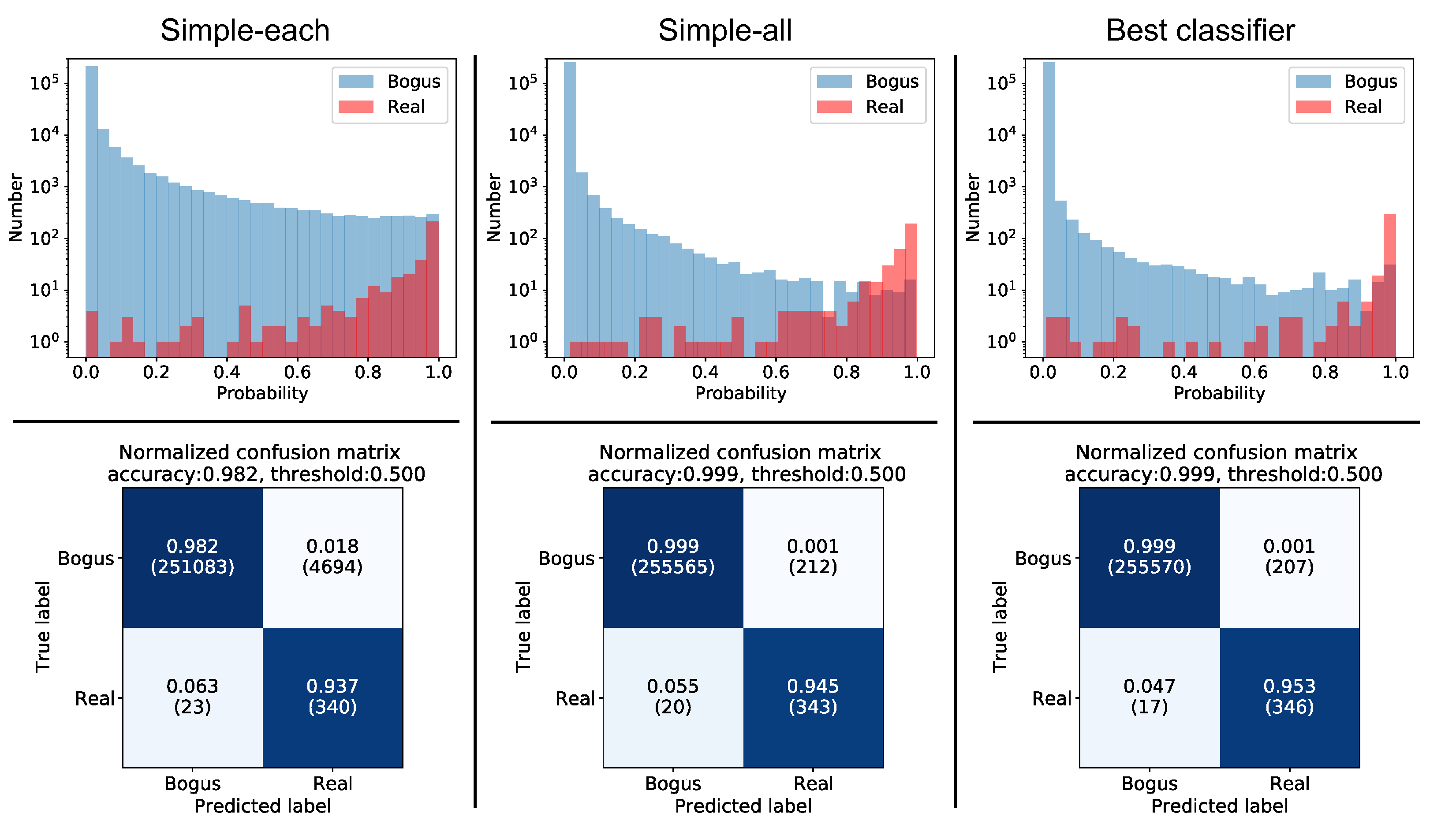}
  \end{center}
  \caption{%
  Upper panels: distribution of output probability (to be Real) for each classifier. Red and blue colors show the histograms for Real and Bogus samples, respectively. Lower panels: Confusion matrix of each classifier. The predicted labels in all the classifiers are the ones with a threshold of 0.5. The ratio in each row is normalized to 1. The numbers in parentheses represent the raw numbers.
  }%
  \label{fig:score_CM}
\end{figure*}

\begin{figure}[Htbp]
  \begin{center}
     \includegraphics[width=\columnwidth]{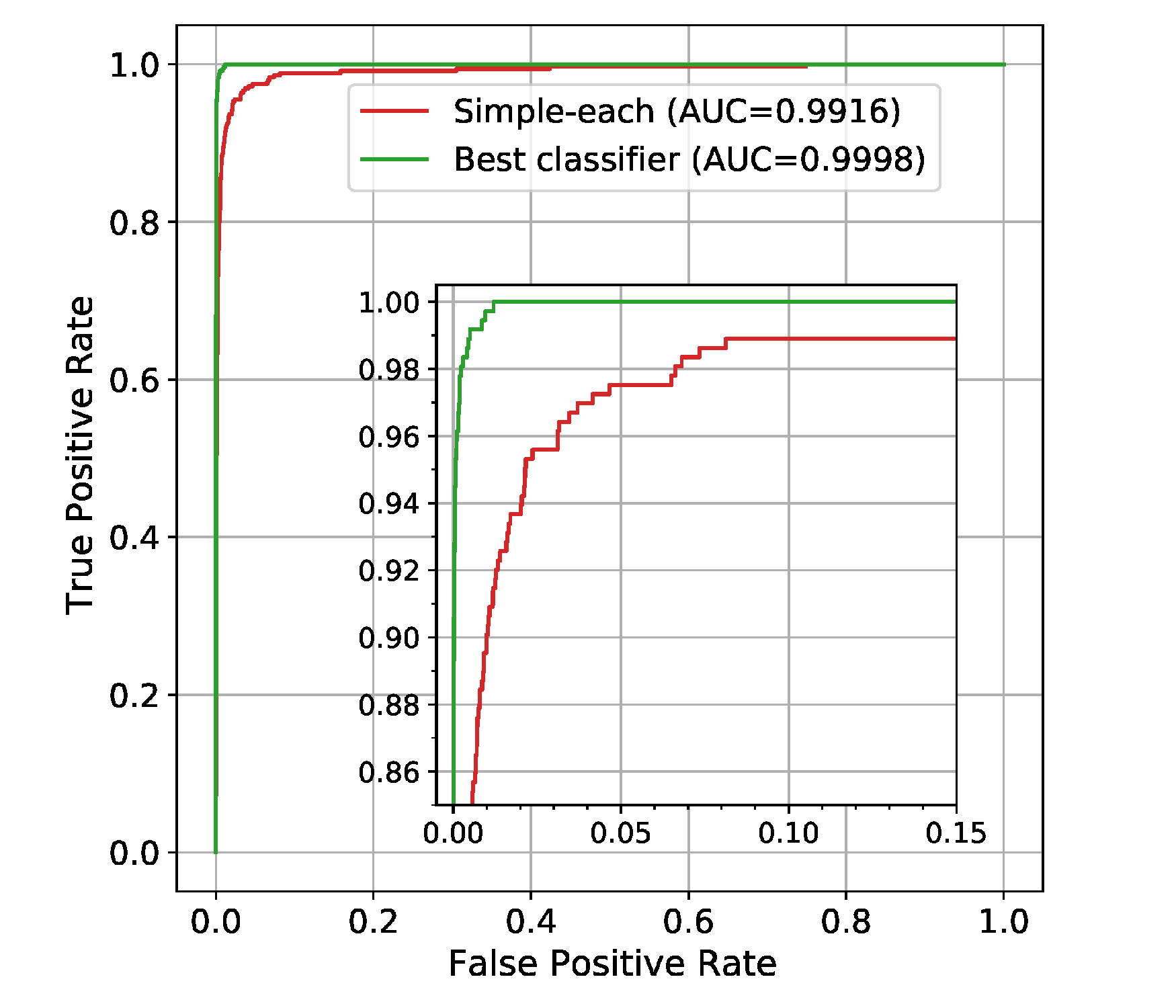}
  \end{center}
  \caption{%
  ROC curves of the Simple-each classifier and the best classifier.
  }%
  \label{fig:ROC}
\end{figure}

\begin{figure}[Htbp]
  \begin{center}
     \includegraphics[width=\columnwidth]{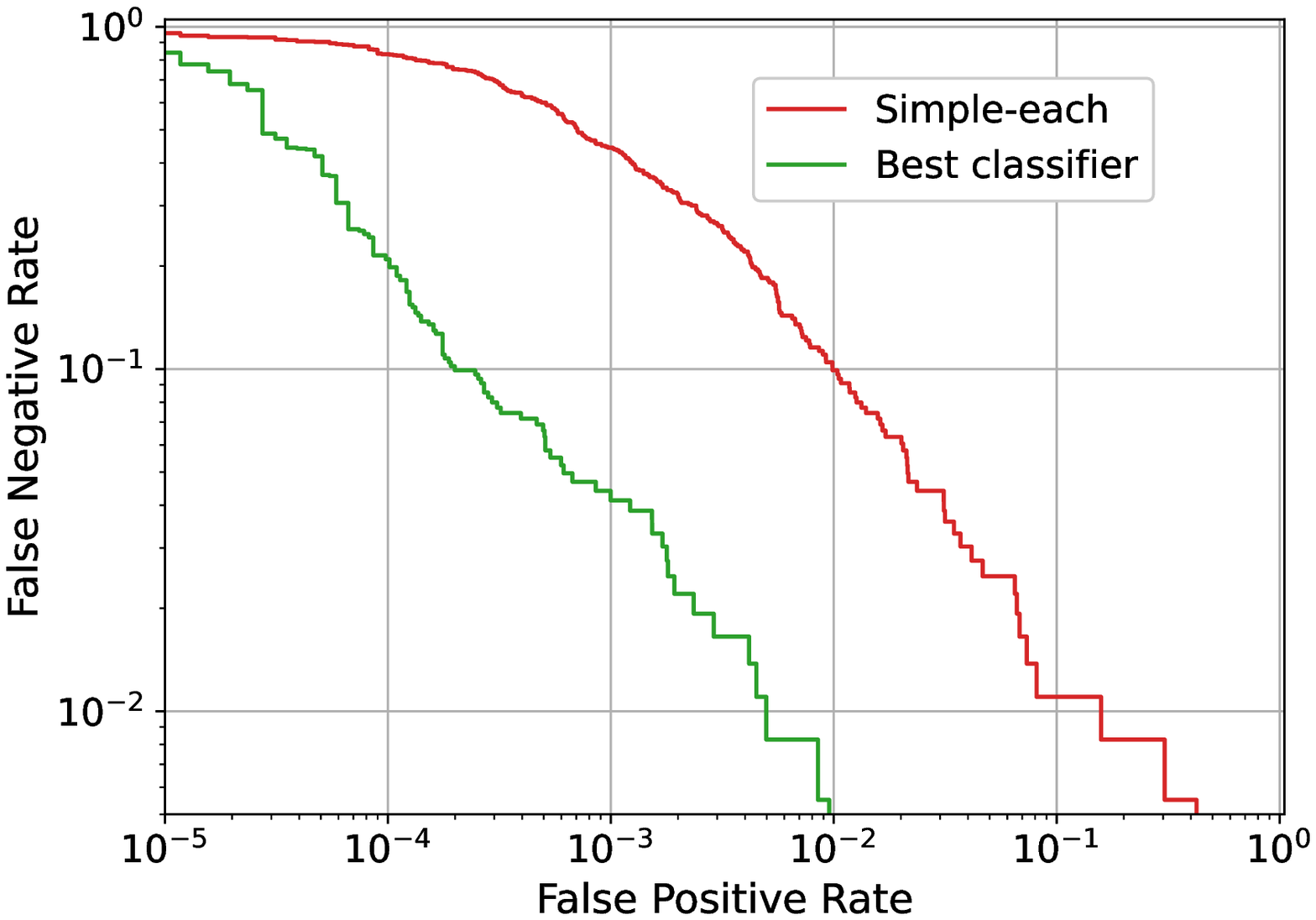}
  \end{center}
  \caption{%
  FNR against FPR for the Simple-each classifier and the best classifier.
  }%
  \label{fig:FPRFNR}
\end{figure}

Finally, we compare the classification performance in all the cases with different objective function combinations and with/without handling label errors.
AUC and FPR for all the cases are summarized in figures\ \ref{fig:auc_complex} and \ref{fig:fpr_complex}.
Comparing all the cases, the case combining the three objective functions and using semi-supervised learning achieved the best results (index 9).
For comparison, the prediction distributions and confusion matrices for the Simple-each case, the Simple-all case, and the best classifier are shown in figure\ \ref{fig:score_CM}.
In the best classifier, the number of Bogus misclassified as Real is further reduced as compared with the Simple-all case, and the false positive of the confusion matrix is 1/23 of that of the Simple-each case.
The ROC curve of the best classifier is shown with the green line in figure\ \ref{fig:ROC}, where the AUC reaches to 0.9998.
Similarly, the relationship between FPR and false negative rate (FNR$=1-$TPR) is plotted as figure\ \ref{fig:FPRFNR}. 
The FPR decreases to 0.0002 when FNR=0.1 (TPR=0.9).

\begin{table*}[Htbp]
\tbl{Summary of classification performance.}{
\begin{tabular}{llllllll|ll}
\hline
Model type & Data type & CE & AUC & VAT & $\lambda_\mathrm{ce}$ & $\lambda_\mathrm{ech}$ & $\lambda_\mathrm{lds}$ & AUC & FPR@TPR=0.9 \\ \hline
simple & each & $\checkmark$ &  &  & 1.0 & 0.0 & 0.0& 0.9916& $9.868\times 10^{-3}$ \\
simple & mix & $\checkmark$ &  &  & 1.0 & 0.0 & 0.0& 0.9964& $5.028\times 10^{-3}$ \\
simple & all & $\checkmark$ &  &  & 1.0 & 0.0 & 0.0& 0.9997& $3.323\times 10^{-4}$ \\
complex & all & $\checkmark$ &  &  & 1.0 & 0.0 & 0.0& 0.9993& $7.272\times 10^{-4}$ \\
complex & removed & $\checkmark$ &  &  & 1.0 & 0.0 & 0.0& 0.9993& $8.367\times 10^{-4}$ \\
complex & all & $\checkmark$ & $\checkmark$ &  & 0.3 & 0.7 & 0.0& 0.9992& $1.306\times 10^{-3}$ \\
complex & removed & $\checkmark$ & $\checkmark$ &  & 0.4 & 0.6 & 0.0& 0.9996& $5.083\times 10^{-4}$ \\
complex & all & $\checkmark$ &  & $\checkmark$ & 0.4 & 0.0 & 0.6& 0.9997& $1.603\times 10^{-4}$ \\
complex & unlabeled & $\checkmark$ &  & $\checkmark$ & 0.6 & 0.0 & 0.4& 0.9997& $1.838\times 10^{-4}$ \\
complex & all & $\checkmark$ & $\checkmark$ & $\checkmark$ & 0.3 & 0.3 & 0.4& 0.9997& $2.541\times 10^{-4}$ \\
complex & unlabeled & $\checkmark$ & $\checkmark$ & $\checkmark$ & 0.9 & 0.05 & 0.05& 0.9998& $1.994\times 10^{-4}$ \\
\hline
\end{tabular}
}\label{tab:performance}
\end{table*}

%

\section{Discussion}\label{sec:discussion}
In this section, we review the performance improvement in our Real/Bogus classification and discuss the key factors of the improvement, as well as the actual performance after implementing the best classifier in the Tomo-e Gozen transient pipeline.

\subsection{Performance analysis}\label{sec:performance}
\begin{figure}[thbp]
  \begin{center}
     \includegraphics[width=\columnwidth]{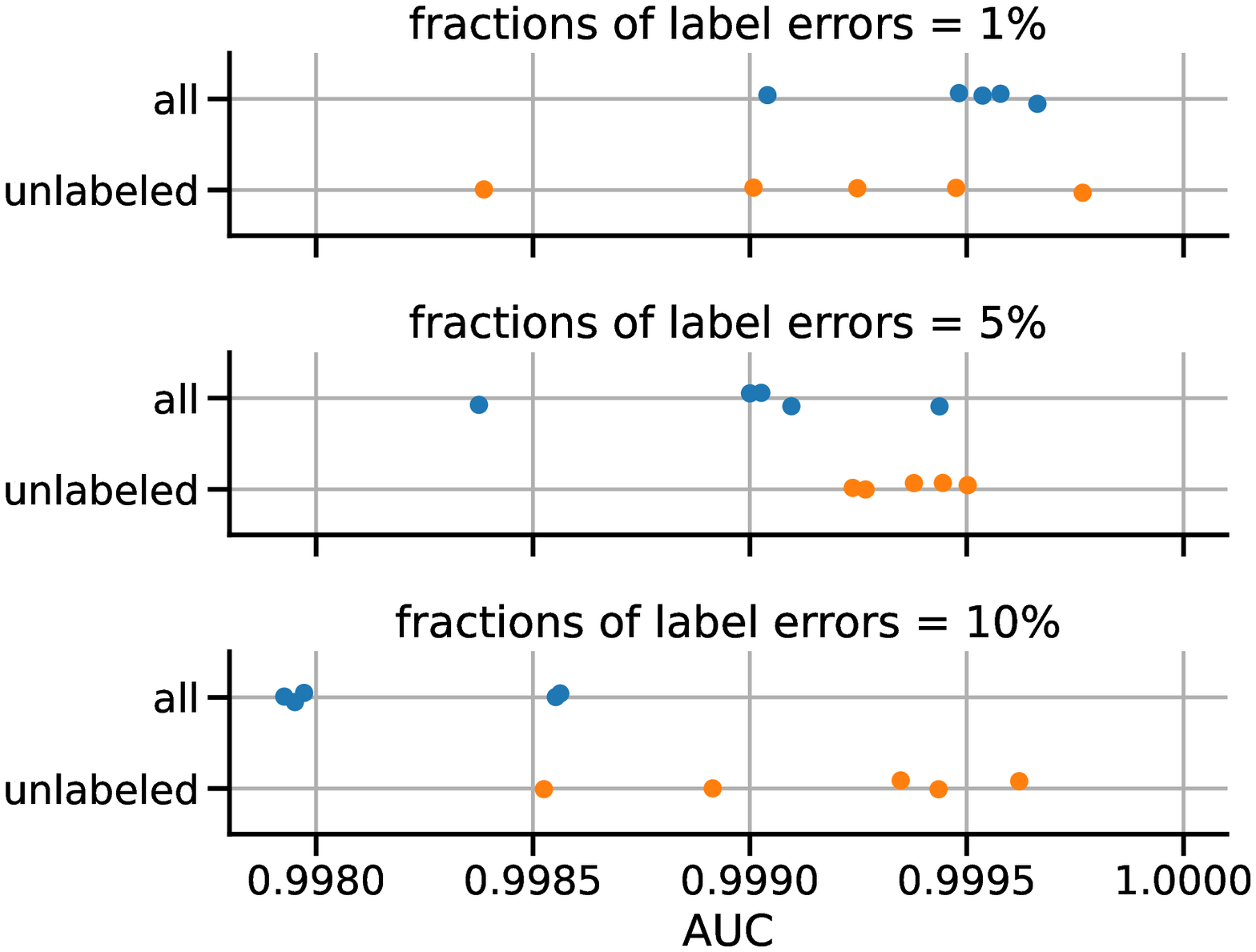}
  \end{center}
  \caption{%
  AUC for different fractions of label errors.
  }%
  \label{fig:auc_noise}
\end{figure}

\begin{figure}[thbp]
  \begin{center}
     \includegraphics[width=\columnwidth]{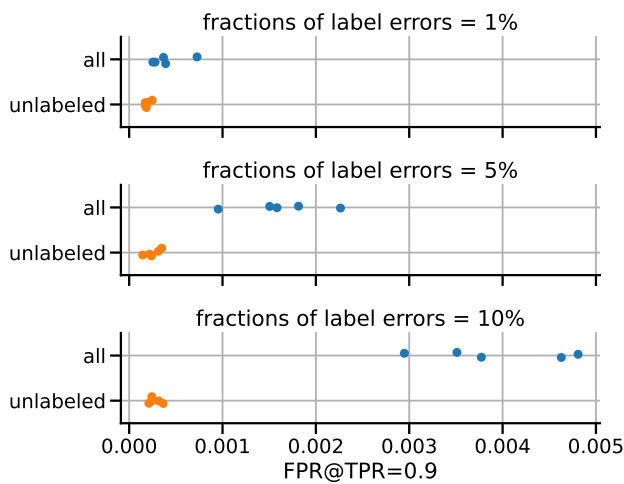}
  \end{center}
  \caption{%
  Same as figure\ \ref{fig:auc_noise} but for FPR at TPR$=$0.9.
  }%
  \label{fig:fpr_noise}
\end{figure}

The following is a summary of the improvement in the Tomo-e Gozen Real/Bogus classification performance.
As shown in table\ \ref{tab:performance}, the best performance is achieved when all of the following conditions are satisfied:
\begin{itemize}
\item Training data for all the sensors are combined.
\item All three objective functions are used.
\item Semi-supervised learning is applied.
\end{itemize}
Compared to the Simple-each case, the AUC improves from 0.9916 to 0.9998 and FPR at TPR$=$0.9 decreases from 0.0099 to 0.0002.
Comparing the performance with the ZTF Real/Bogus classifier (FPR$=$0.017 at TPR$=$0.983, \cite{2019MNRAS.489.3582D}),
our best classifier gives FPR$=$0.003 at the same TPR.
Although exact comparisons cannot be made between Tomo-e Gozen and ZTF because the instruments, pipelines, and Real/Bogus ratio are different, we achieve FPR comparable to that of the ZTF classifier.
We then discuss key factors in improving classification performance.
First, combining training data for different sensors significantly improved performance.
This improvement is a result of the data augmentation effect by mixing data from multiple sensors with different characteristics. Also, the increasing the number of training data improves the performance, as expected.
In addition, incorporating LDS loss into the objective function improves classification performance with the Complex model compared to the Simple-all case.
The Complex model tends to overfit the training data because it can handle complex representations, resulting in lower performance on the validation data.
On the other hand, when LDS loss is included in the objective function, the LDS-based regularization avoids overfitting and improves performance.
We also show that, by setting the label error samples to unlabeled samples and by performing semi-supervised learning, the performance is further improved.
We here investigate whether the proposed method also works when there are more label errors in training data.
The fraction of label errors in our original training dataset is about 1\%. We artificially increase the fraction of label errors by inverting the labels based on the estimated fraction of label errors (figure\ \ref{fig:mislabel}).
We then classify the data with our best method, semi-supervised learning with data containing unlabeled samples, and compare the results with those obtained by training with label errors remaining.
Figures\ \ref{fig:auc_noise} and \ref{fig:fpr_noise} compare the AUC and FPR when the ratio of label errors is increased to about 5\% and 10\%.
It is found that the degree of improvement is higher as the fraction of label errors increases.
When the fraction of label errors is 1\%, by handling label errors, the average FPR decreases from 0.0004 to 0.0002, i.e., an improvement by a factor of about 2.
On the other hand, when the fraction of label errors is 10\%, the average FPR decreases from 0.0039 to 0.0003, which corresponds to the improvement by a factor of 13.
This means that our method is more effective for datasets with higher fractions of label errors.

\subsection{Performance in actual operations}\label{sec:actualope}
\begin{figure}[thbp]
  \begin{center}
     \includegraphics[width=\columnwidth]{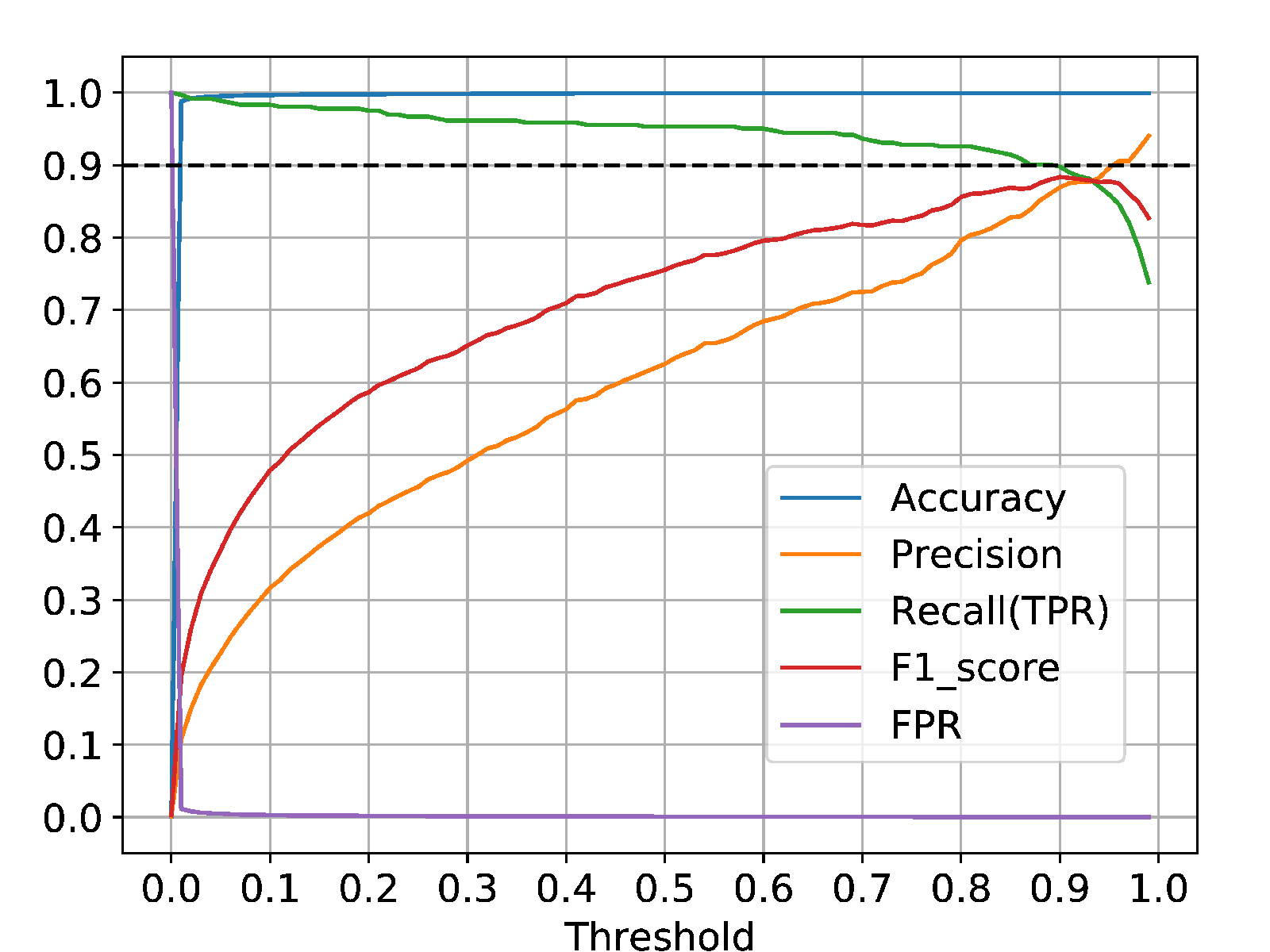}
  \end{center}
  \caption{%
  Variation of each metric as a function of threshold for the best classifier.
  }%
  \label{fig:metrics}
\end{figure}

\begin{figure*}[thbp]
  \begin{center}
     \includegraphics[width=2\columnwidth]{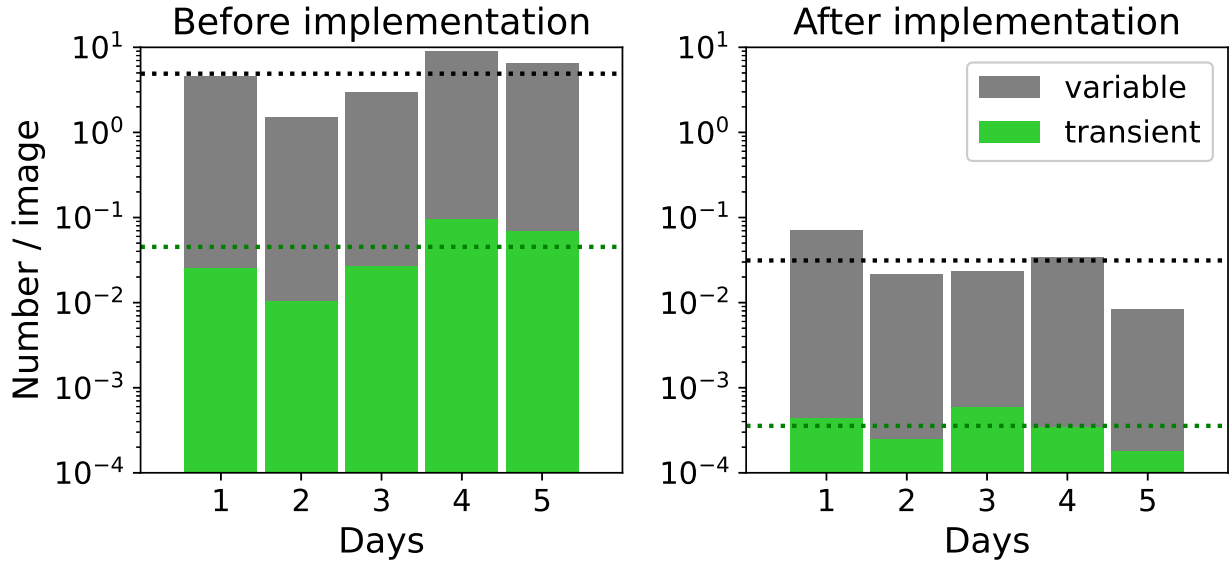}
  \end{center}
  \caption{%
  The number of objects registered in the Tomo-e transient database before (left panel) and after (right panel) implementation of the best classifier. The conventional classifier is trained in the same way as the Simple-each classifier. Horizontal dotted lines show five-day averages.
  }%
  \label{fig:registered}
\end{figure*}

Finally, we discuss the actual performance of our best classifier when implemented in the data analysis pipeline of Tomo-e Gozen.
Prior to implementation, we determined the threshold for the classification in the actual operation.
Figure\ \ref{fig:metrics} shows the variation of each metric as a function of the threshold.
We set the threshold score to be 0.85, which gives the best precision while keeping TPR above 0.9.
We investigate the changes in the number of transient candidates registered in the Tomo-e Gozen transient database before and after the implementation.
Figure\ \ref{fig:registered} shows the number of registrations to the database before and after implementation over a five-day period.
As a rule for registration in the database, objects detected for the first time by the classifier are registered as ``variable''.
Among the variable candidates, those detected twice at the same coordinates are registered as ``transient''.
After the implementation of the new classifier, the average numbers of variable and transient registrations were reduced to 1/160 and 1/130, respectively.
To confirm that the new classifier does not miss real objects, we examined the recovery rate of real objects.
By matching registered transients with TNS objects, we confirmed that the ratio of the number of matches to the number of TNS objects is comparable before and after implementation.
Furthermore, the fraction of the TNS object rate among the registered transient candidates is 86 times higher after the implementation.
This indicates that Bogus, which are incorrectly registered due to misclassification, are greatly reduced.
The number of final transient candidates are reduced from about about 5,000 to 40 objects per day.
This rate is a level at which human can check visually, and enables effective target selection for follow-up observations in a short time. 

\section{Conclusions}\label{sec:conclusion}
In this paper, we have presented a new Real/Bogus classification scheme by handling label errors in the training data for the Tomo-e Gozen transient survey.
In the wide-field, high-frequency survey with Tomo-e Gozen, which mainly targets early supernovae and rapid transients, the performance of conventional CNN classifiers was not sufficient to extract follow-up targets, because the number of Bogus was an order of $10^3$ per day.
Therefore, we developed a two-step training method: (1) normal supervised learning to detect label errors in the training data, and (2) 
semi-supervised learning with training data which include potential label errors as unlabeled samples.

The best classifier with this method achieves an AUC of 0.9998 and FPR of 0.0002 at TPR$=$0.9 for validation data prepared from actual observations.
Our training method does not require human effort to relabel the samples with potential label errors.
We also show that our method gives a higher performance improvement when the fraction of label errors is higher.
Finally, we implemented the developed classifier in the Tomo-e Gozen pipeline. After implementation, the number of registered transient candidates was reduced by a factor of about 100, to 40 candidates per day, while maintaining the recovery rate of real transients.
This enables more efficient selection of follow-up targets.

\begin{ack}
We thank Yasuhiro Imoto for his significant contributions to the development of the classifiers.
We are grateful to the anonymous referee for insightful suggestions.

This work has been supported by Japan Science and Technology Agency (JST)  AIP Acceleration Research Grant Number JP20317829 and the Japan Society for the Promotion of Science (JSPS) KAKENHI grants 21H04491, 18H05223, and 17H06363.

This work is supported in part by the Optical and Near-Infrared Astronomy Inter-University Cooperation Program.

The Pan-STARRS1 Surveys (PS1) and the PS1 public science archive have been made possible through contributions by the Institute for Astronomy, the University of Hawaii, the Pan-STARRS Project Office, the Max-Planck Society and its participating institutes, the Max Planck Institute for Astronomy, Heidelberg and the Max Planck Institute for Extraterrestrial Physics, Garching, The Johns Hopkins University, Durham University, the University of Edinburgh, the Queen's University Belfast, the Harvard-Smithsonian Center for Astrophysics, the Las Cumbres Observatory Global Telescope Network Incorporated, the National Central University of Taiwan, the Space Telescope Science Institute, the National Aeronautics and Space Administration under Grant No. NNX08AR22G issued through the Planetary Science Division of the NASA Science Mission Directorate, the National Science Foundation Grant No. AST-1238877, the University of Maryland, Eotvos Lorand University (ELTE), the Los Alamos National Laboratory, and the Gordon and Betty Moore Foundation.
\end{ack}

\appendix 
\section*{Details of objective function}
We here describe the details of the objective functions proposed in section\ \ref{sec:objective_functions}.

\paragraph*{Cross-entropy}:
The cross-entropy function is an objective function commonly used in training classifiers.
It works to match the output of the neural network to the teacher label for each sample.
The cross-entropy function is as follows:
\begin{eqnarray}
    \lefteqn{L_\mathrm{ce} \left( \left\{\mathbf{x}_l, y\right\} ; \boldsymbol{\theta} \right)} \nonumber \\
    & = \frac{1}{N_l} \sum_{i=1}^{N_l} &\left\{- I\left(y_i=c^+\right) \log p\left(y=c^+\mid \mathbf{x}_i ;\boldsymbol{\theta}\right) \right. \nonumber \\
    & & \left. \; - I\left(y_i=c^-\right) \log p\left(y=c^-\mid \mathbf{x}_i ;\boldsymbol{\theta}\right) \right\},
\end{eqnarray}
\begin{eqnarray}
    \lefteqn{p\left(y=c\mid \mathbf{x};\boldsymbol{\theta}\right)} \nonumber \\
    & = & \frac{\exp\left(f\left(y=c\mid \mathbf{x}; \boldsymbol{\theta}\right)\right)}{\exp\left(f\left(y=c^+\mid \mathbf{x}; \boldsymbol{\theta}\right)\right) + \exp\left(f\left(y=c^-\mid \mathbf{x}; \boldsymbol{\theta}\right)\right)}
    .
\end{eqnarray}
Here $N_l$ is the number of labeled samples, $I\left(\right)$ is the indicator function, $c^+$ and $c^-$ are the labels of positive and negative examples, respectively, and $f\left(y=c\mid \mathbf{x};\boldsymbol{\theta}\right)$ is the output value corresponding to the label $c$ ($\mathbf{x}$ is input to the neural network).

\paragraph*{exp-Cross-hinge loss (AUC maximization)}:
If the sample ratio is highly biased, the apparent performance could be improved by always predicting the dominant class regardless of the input data.
For example, in transient surveys, Bogus objects are always dominant over Real objects.
In such a case, a classifier that classifies all the input to be Bogus can achieve a high score.
However, obviously such a classifier is not useful to detect real transients.
Approaches to handle imbalanced dataset include down-sampling of the majority class (e.g., \cite{2019MNRAS.488.4858H}) and giving extra weight to the minority class (e.g., \cite{2021AJ....161..267V}).
In our proposed method, we handle imbalanced data by incorporating exp-Cross-hinge loss (\cite{misc/27342721}) into the objective function, as described below.
The exp-Cross-hinge loss is a loss function for pairs of positive and negative examples in the dataset.
When the score of the negative sample becomes larger than the score of the positive sample, a loss corresponding to the difference occurs.
In contrast, when the score of the positive example is larger than the score of the negative example, the hinge function prevents the loss of the set from falling below a certain level.
In addition, expanding the difference between the negative and positive scores with the exponential function enables learning even when the difference is small.
For all pairs of positive and negative samples, this loss function is minimized when the score of the positive sample is greater than the score of the negative sample including the margin.
The definition is as follows:
\begin{eqnarray}
    \lefteqn{L_\mathrm{ech} \left( \left\{\mathbf{x}^+\right\}, \left\{\mathbf{x}^-\right\}; \boldsymbol{\theta} \right)} \nonumber \\
    & = & \frac{1}{N^+ N^-} \sum_{i=1}^{N^+} \sum_{j=1}^{N^-} \exp \left\{ \left[ s\left(\mathbf{x}_j^-; \boldsymbol{\theta}\right) - s\left(\mathbf{x}_i^+; \boldsymbol{\theta}\right) + \xi \right]_+ \right\}
    .
\end{eqnarray}
Here, $s\left(\right)$ is the score function parameterized by $\boldsymbol{\theta}$, $\mathbf{x}^+$ and $\mathbf{x}^-$ are the positive and negative samples, $N^+$ and $N^-$ are the numbers of positive and negative samples, respectively, and $\xi$ is the margin between the positive and negative sample scores.
In the equation, $\left[\bullet\right]_+$ is the hinge function, defined as 
\begin{eqnarray}
    \left[z\right]_+ = \left\{
\begin{array}{ll}
z & (z \geq 0)\\
0 & (z < 0)
\end{array}
\right.
.
\end{eqnarray}
We define the score function using the outputs of the neural network as follows:
\begin{eqnarray}
    \label{eq:score}
    s\left(\mathbf{x}; \boldsymbol{\theta}\right) = f\left(y=c^+\mid \mathbf{x}; \boldsymbol{\theta}\right) - f\left(y=c^-\mid \mathbf{x}; \boldsymbol{\theta}\right) .
\end{eqnarray}
The AUC is a measure for the performance of a binary classifier when we want to maximize the true positive rate and minimize the false positive rate.
AUC is generally defined as the area under the ROC curve.
Alternatively, it can be defined as the ratio of pairs in which the score of the positive sample is greater than the score of the negative sample among all pairs of positive and negative samples in the dataset.
Since AUC is a discontinuous function, it is difficult to maximize AUC directly.
On the other hand, since the exp-Cross-hinge function is a relaxation of the AUC function to a continuous function, we expect to obtain an approximate solution for AUC maximization by minimizing the exp-Cross-hinge function.
\paragraph*{Virtual Adversarial Training (VAT)}:
We perform the Virtual Adversarial Training (VAT) (\cite{miyato2016distributional}).
The VAT is a training method with the local distributional smoothness (LDS) as a regularization term.
The LDS is a new notion of smoothness for the outputs of models.
In VAT, special perturbations that maximize the changes of the outputs of the neural network are added to input images.
The neural network is trained to minimize the change of the outputs, thus smoothing regularization.
Therefore, it is expected to be robust to the influence of input perturbation.
The objective function is as follows:
\begin{eqnarray}
    \lefteqn{L_\mathrm{lds} \left(\left\{ \mathbf{x}_l \right\} \cup \left\{ \mathbf{x}_u \right\}; \boldsymbol{\theta}\right)} \nonumber \\
    & = & \frac{1}{N_l + N_u} \sum_{i=1}^{N_l + N_u} \mathrm{KL} \left[ p\left(y\mid \mathbf{x}_i; \boldsymbol{\theta}\right) \middle| \middle| p\left(y \mid \mathbf{x}_i  + \mathbf{r}_i; \boldsymbol{\theta}\right) \right]
    ,
\end{eqnarray}
where $N_l$ and $N_u$ are the numbers of labeled samples and unlabeled samples, respectively, $\mathrm{KL} \left[ \bullet \middle| \middle| \bullet \right]$ is KL divergence which is a measure of the difference between two distributions, and $\mathbf{r}_i$ is the virtual adversarial perturbation of the $i$th sample.
Here, $\mathbf{r}_i$ is a tiny perturbation that maximizes the change in the classifier output and it is defined as
\begin{eqnarray}
    \mathbf{r}_i
    = \argmax_{\mathbf{r}, \; \mathrm{w.r.t.} \left\|\mathbf{r}\right\|^2<\epsilon} \mathrm{KL} \left[ p\left(y_i\mid \mathbf{x}_i; \boldsymbol{\theta}\right) \middle| \middle| p\left(y_i \mid \mathbf{x}_i  + \mathbf{r}; \boldsymbol{\theta}\right) \right],
\end{eqnarray}
where $\epsilon$ is the size of the perturbation.
It is not practical to obtain $\mathbf{r}_i$ exactly because it is computationally expensive.
Instead, we efficiently obtain the approximate virtual adversarial perturbation by using the method shown in algorithm 1 of \citet{miyato2016distributional}.
Because no label information is needed to compute $\mathbf{r}_i$, the VAT objective function can be used even for samples without labels.

%
%
 \newcommand{\noop}[1]{}

\end{document}